\documentclass[aps,prb,twocolumn,showpacs,amsmath,tightenlines,
superscriptaddress]{revtex4} 

\usepackage{graphicx, bm}

\begin{document}

\title{AC- and DC-driven noise and $I$-$V$ characteristics of magnetic
nanostructures}

\author{O. A. Tretiakov}

\affiliation{Department of Physics, New York University, New York, New
York 10003, USA}

\author{Aditi Mitra}

\affiliation{Department of Physics, New York University, New York, New
York 10003, USA}

\date{November 12, 2009}

\begin{abstract}
We study a structure consisting of a ferromagnetic (F) layer coupled
to two normal metal (N) leads. The system is driven out of equilibrium
by the simultaneous application of external dc and ac voltages across
the N/F/N structure. Using the Keldysh diagrammatic approach, and
modeling the ferromagnet as a classical spin of size $\mathbf{S} \gg 1$,
we derive the Langevin equation for the magnetization dynamics and
calculate the noise correlator. We find that the noise has an explicit
frequency dependence in addition to depending on the characteristics
of the ac and dc drive.  Further, we calculate the current-voltage
characteristics of the structure to ${\cal O}\left(1/S^2\right)$ and
find that the nonequilibrium dynamics of the ferromagnetic layer gives
rise to corrections to the current that are both linear and nonlinear
in voltage.
\end{abstract}


\maketitle

\section{Introduction}

Magnetization dynamics in small nanomagnets has recently attracted a
lot of theoretical \cite{Kamenev08, Bauer09, Mitra06, Duine07,
Vavilov09} and experimental \cite{magnet_dyn_exp, Tserkovnyak:review}
attention due to advances in manufacturing magnetic
nanostructures. The topic of magnetization noise has become an
exciting subject owing to its possible influence on magnetization
switching \cite{Bauer09} and conductivity of these
structures.\cite{Brataas07}  It has been shown that the noise in
magnetic structures, such as spin valves, can be colored, i.e., it can
have a nontrivial frequency dependence.\cite{Xiao09}  In diffusive
metallic conductors colored noise has been observed
experimentally,\cite{Prober} however in magnetic structures it still
requires further investigation.

Experiments involving magnetic nanostructures typically involve the
simultaneous application of dc and ac voltage where the ac bias is
found to aid the magnetization switching.  Therefore in this paper we
study a normal metal/ferromagnet/normal metal (N/F/N) structure which
has been driven out of equilibrium by the simultaneous application of
a dc and ac voltage.  We show that the effect of this driving is to
produce a noise in the magnetization dynamics that is colored. In
addition we determine how the $I$-$V$ characteristics of the device
are affected by the dynamics of the ferromagnetic layer.

\begin{figure}
\includegraphics[width=0.98\columnwidth]{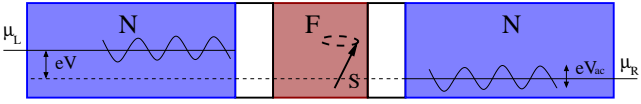}
\caption{(Color online) A sketch of the N/F/N structure. The left and
  right normal-metal leads are coupled through tunnel barriers to the
  ferromagnetic layer.}
\label{fig-structure}
\end{figure}

The schematic of the N/F/N structure we study is shown in
Fig.~\ref{fig-structure}.  The ferromagnetic layer is assumed to be
small so that it may be modeled as a single-domain magnet.  At the
same time, the spin of the magnet is considered to be large (spin
$\mathbf{S}\gg1$) so that it can be treated as a classical variable.
The aim of this paper is twofold, one is to derive the Langevin
equation for the magnetization dynamics, and second is to present a
calculation of the $I$-$V$ characteristics. In the absence of any
magnetization dynamics, the N/F/N structure is Ohmic.\cite{1} We show
that the dynamics of the ferromagnet gives rise to corrections to the
$I$-$V$ characteristics that are both linear and nonlinear in voltage.

The paper is organized as follows. In section~\ref{model} we present
the model. In section~\ref{secI} we study the nonequilibrium
properties of the model in the limit $S \rightarrow \infty$, when the
magnetization is static.  In section~\ref{secII} we study small
fluctuations of the magnetization about the ordering direction thus
deriving the Langevin equation and the noise spectrum.  The results of
this section are then used in section~\ref{secIII} to calculate the
corrections to the current-voltage characteristics arising due to the
magnetization fluctuations. Finally in section~\ref{secIV} we
summarize our results.

\section{Model} \label{model}
  
We consider a model Hamiltonian $H=H_{m}+H_{l}+H_{t}$, where $H_{m}$
describes the ferromagnetic layer, $H_{l}$ represents the two
normal-metal leads, and $H_{t}$ models the tunneling between the leads
and magnetic layer.  The Hamiltonian for the magnetic layer $H_{m}$ is
\begin{equation}
H_{m} =-(DS_{z}^{2}+BS_{z})+J\sum_{i}\mathbf{S}\cdot\mathbf{s}_{i}
+\sum_{\mathbf{k}\sigma}\epsilon_{\mathbf{k}}^{d}
d_{\mathbf{k}\sigma}^{\dagger}d_{\mathbf{k}\sigma}.
\label{H_m}
\end{equation}
Here the first term models a material (or shape) anisotropy with the
anisotropy constant $D$, the second term describes the interaction of
the macrospin $\mathbf{S}$ with the magnetic field $\mathbf{B}$
applied for simplicity in the same $\mathbf{z}$-direction as the
anisotropy. The third term describes the interaction of the macrospin
with the spins of the itinerant electrons $\mathbf{s}_{i}$ as in the
$s$-$d$ model.  It can be rewritten as
$J\sum_{i}(S_{z}s_{zi}+S_{+}s_{-i} +S_{-}s_{+i})$, where
$S_{\pm}=(S_{x}\pm iS_{y})/2$, $s_{\pm i}=s_{xi}\pm is_{yi}$, and
$\mathbf{s}_{i}=\frac{1}{2}
\sum_{\mathbf{k},\sigma,\alpha,\beta}d_{\mathbf{k}\sigma,\alpha}^{\dagger}
\mbox{\boldmath$\sigma$}_{\alpha\beta}d_{\mathbf{k}\sigma,\beta}$ with
$\sigma_{\alpha\beta}$ being the components of Pauli matrices. Here
$d_{\mathbf{k}\sigma}^{\dagger}(d_{\mathbf{k}\sigma})$ creates
(destroys) an electron in state with momentum $\mathbf{k}$ and spin
component $\sigma$.  The pure macrospin part of the Hamiltonian can be
rewritten as $-(DS_{z}^{2}+BS_{z}) \simeq \text{const}+bS_{+}S_{-}$,
where the constant part is $-DS^{2}-BS$ and $b=4(D+B/(2S))$.
Nanomagnets are typically characterized by a significant anisotropy.
This along with the fact that ${\bf S} \gg 1$ implies that the
fluctuations of the nanomagnet about the ordering direction are
small. Our theoretical treatment will therefore involve a perturbative
expansion in spin fluctuations, which as we shall show is equivalent
to an expansion in $1/S$.

We assume that the electrons in the leads are non-interacting. To
model the ac bias voltage we introduce a time-dependence of the lead
single-particle energies \cite{Wingreen} namely,
$\epsilon_{k\alpha}(t) =\epsilon_{k\alpha}^{0}
+V_{ac}\cos(\omega_{0}t+\varphi_{\alpha})$ where $\alpha$ labels the
left ($L$) or right ($R$) lead, and $V_{ac},\omega_0$ are,
respectively, the amplitude and frequency of the ac bias.  Thus, the
lead Hamiltonian is
\begin{equation}
H_{l}=\sum_{k\sigma,\alpha\in L,R} 
\epsilon_{k\alpha}(t)c_{k\sigma\alpha}^{\dagger}c_{k\sigma\alpha} 
\label{eq:Hl}\;,
\end{equation}
The coupling between the leads and magnetic layer is
\begin{equation}
H_{t} = \sum_{\mathbf{k},k_{0},\sigma,\alpha\in L,R}
(t_{\alpha}c_{\mathbf{k}k_{0}\sigma\alpha}^{\dagger}d_{\mathbf{k}\sigma}
+\mathrm{H.c.}).
\label{H_t}
\end{equation}
In Eq.~(\ref{eq:Hl}) $k=(\mathbf{k},k_{0})$ where $\mathbf{k}$ is a
two-dimensional momentum in the plane perpendicular to the tunneling
direction and is assumed to be conserved on tunneling.

To study this nonequilibrium problem we employ the Keldysh
formalism.\cite{Keldysh65, Mitra06} We introduce variables
$\mathbf{S}^{cl}=(\mathbf{S}^{+} + \mathbf{S}^{-})/2$ and
$\mathbf{S}^{q} = (\mathbf{S}^{-} -\mathbf{S}^{+})/2$ where the upper
$\pm$ indices correspond to the time-ordered (anti-time-ordered)
directions on the Keldysh contour, and the Keldysh path integral takes
the form, $\mathcal{Z}_{K}=\int \mathcal{D} [\mathbf{S}^{cl},
  \mathbf{S}^{q}]\, e^{-i\mathcal{S}_{K}}$. Here $\mathcal{S}_{K}$ is
the effective action for the macrospin obtained formally by
integrating out all fermionic degrees of freedom:
\begin{eqnarray}
\!\!\!\!\!\mathcal{S}_{K}\!\! &=&\!\! 2b\textrm{Tr} \left(S_{+}^{cl}S_{-}^{q}
+S_{+}^{q}S_{-}^{cl}\right)
\nonumber \\
&+&\!\! i\textrm{Tr}\ln\left[\hat{g}_{d\sigma}^{-1}-\hat{\Sigma}
-J\hat{S}_{z}\frac{\sigma_{z}}{2}
-J\hat{S}_{+}\sigma_{-}-J\hat{S}_{-}\sigma_{+}\right]\! ,
\label{eq:eff_action}
\end{eqnarray}
where $\hat{S}_{a=z,\pm} = \left(\begin{array}{cc} S_{a}^{cl} &
S_{a}^{q}\\ S_{a}^{q} & S_{a}^{cl}\end{array}\right)$ and
$\hat{g}_{d\sigma}=\left(\begin{array}{cc} g_{d\sigma}^{R} &
g_{d\sigma}^{K}\\ 0 & g_{d\sigma}^{A}\end{array}\right)$ are the
Green's functions of the free electrons in the magnetic layer,
$\sigma_z$ and $\sigma_{\pm}=(\sigma_x \pm \sigma_y)/2$ are Pauli
matrices, and $\hat{\Sigma}=\left(\begin{array}{cc} \Sigma^{R} &
\Sigma^{K}\\ 0 & \Sigma^{A}\end{array}\right)$ is the self-energy due
to coupling to the leads. As we shall show, $\hat{\Sigma}$ depends on
the ac and dc bias and is independent of $\sigma$ because the leads
are non-magnetic.  In what follows we make the assumption that the
fluctuations of the macrospin from the ordering direction are
small. Thus we write $J \hat{S}_z = JS + J \delta \hat{S}_z$, and
eventually expand Eq.~(\ref{eq:eff_action}) perturbatively in the
fluctuations $\delta \hat{{S}}_z, \hat{S}_{\pm}$.

\section{Green's functions in the magnetic layer} \label{secI}

We first discuss the properties of the nonequilibrium system when the
magnetization does not fluctuate. Denoting $G_{0}$ to be the Green's
function of the electrons in the magnetic layer when $\delta
\hat{{S}}_z= \hat{S}_{\pm}=0$, Eq.~(\ref{eq:eff_action}) implies
\begin{subequations}
\label{eq:G_R_0}
\begin{eqnarray}
\left[G_{0\sigma}^{R}\right]^{-1} &=& \left[g_{d\sigma}^{R}\right]^{-1}
- \Sigma^R - J S\frac{\sigma}{2},
\label{GRdyson}\\
G_{0\sigma}^{K} &=& G_{0\sigma}^{R}\circ\Sigma^{K}\circ G_{0\sigma}^{A}.
\label{GKdyson}
\end{eqnarray}
\end{subequations}
The symbol $\circ$ in Eq.~(\ref{eq:G_R_0}) denotes convolution in the
time domain, and the self-energies due to coupling to leads are
\begin{equation}
\Sigma^{R(K)}(t,t')=\sum_{k_{0},\alpha}t_{\alpha}^{2}
g_{k\sigma\alpha}^{R(K)}(t,t').
\label{eq:sigma}
\end{equation}
$g_{k\sigma\alpha}^{R(K)}$ are the retarded and Keldysh components of
the electron Green's function in the leads and are defined as
\begin{subequations}
\begin{eqnarray}
g_{k\sigma \alpha}^{R}(t,t') &=& -i\theta(t-t')
\left\langle \left\{c_{k\sigma\alpha}(t),c_{k\sigma\alpha}^{\dagger}(t')
\right\}\right\rangle, \\
g_{k\sigma\alpha}^{K}(t,t') & = & -i\left\langle\left[
c_{k\sigma\alpha}(t),c_{k\sigma\alpha}^{\dagger}(t')\right]\right\rangle ,
\end{eqnarray}
\end{subequations}
Since $c_{k\sigma\alpha}(t) =e^{-i\int_{-\infty}^{t}dt_{1}
\epsilon_{k\alpha}(t_{1})} c_{k\sigma\alpha}(-\infty)$, we find
\begin{subequations}
\begin{eqnarray}
&&\!\!\!\!\!\!\!\!\!\!\! g_{k\sigma\alpha}^{R}(t,t')=
-i\theta(t-t')e^{-i\int_{t'}^{t}dt_{1}\epsilon_{k\alpha}(t_{1})},
\label{eq:retarded_lead}\\
&&\!\!\!\!\!\!\!\!\!\!\! g_{k\sigma\alpha}^{K}(t,t') =
-i[1-2f(\epsilon_{k\alpha}^{0}
-\mu_{\alpha})]e^{-i\int_{t'}^{t}dt_{1}
\epsilon_{k\alpha}(t_{1})},
\end{eqnarray}
\end{subequations}
where $f$ is the Fermi distribution function in the leads and we have
used that $\langle [c_{k\sigma\alpha}(-\infty),
c_{k\sigma\alpha}^{\dagger}(-\infty)]\rangle
=1-2f(\epsilon_{k\alpha}^{0}-\mu_{\alpha})$ where $\mu_{\alpha}$ is
the chemical potential of lead $\alpha$ and a dc bias corresponds to
$\mu_L \neq \mu_R$.  Using the identity
$e^{\frac{z}{2}\left(a-\frac{1}{a}\right)}
=\sum_{n=-\infty}^{\infty}a^{n}J_{n}(z)$, where $J_{n}(z)$ are Bessel
functions of the first kind, we find
\begin{eqnarray}
&&\!\!\!\!\!\!\!\!\!\!\!\! g_{k\sigma\alpha}^{R}(t,t')
=-i\theta(t-t')e^{-i\epsilon_{k_{0}\mathbf{k}\alpha}^{0}(t-t')}
\nonumber \\
&&\!\!\!\!\!\!\!\!\!\!\!\!\times\!\left[\sum_{n=-\infty}^{\infty}J_{n}^{2}
\left(\frac{V_{ac}}{\omega_{0}}\right)e^{-in\omega_{0}(t-t')}
\right.
\nonumber \\
&&\!\!\!\!\!\!\!\!\!\!\!\!\left.
+\!\sum_{n\neq m}J_{n}\left(\frac{V_{ac}}{\omega_{0}}\right)
J_{m}\left(\frac{V_{ac}}{\omega_{0}}\right)
e^{i\varphi_{\alpha}(m-n)-i\omega_{0}(mt'-nt)}\right]\!\! .
\label{eq:gkl}
\end{eqnarray}
Changing variables to $\tau=t-t'$ and $T=(t+t')/2$ one may write
$mt'-nt=(n-m)T+(n+m)\tau/2$. In what follows we average Green's
functions over time $T\gg\omega_{0}^{-1}$, where the averaging is
denoted by $\overline{g_{k\alpha}(t,t')}$. This is justified when the
magnetization dynamics is slow compared to $\omega_{0}^{-1}$ (a
precise condition for this will be given in Sec.~\ref{secII}).

As a result of the time averaging, terms corresponding to $n \neq m$ in
Eq.~(\ref{eq:gkl}) vanish.  This leads to a time-averaged retarded
self-energy $\overline{\Sigma^{R}}(\omega) =-i\sum_{\alpha\in
L,R}\Gamma_{\alpha}$ where $\Gamma_{\alpha}=\pi\nu t_{\alpha}^{2}$ is
the decay rate of the electrons into the leads, and $\nu$ is the
density of states in the leads. In what follows we will assume
$\Gamma_{\alpha}$ to be independent of energy.  From
Eq.~(\ref{eq:sigma}), the time-averaged Keldysh component of the
self-energy becomes
\begin{equation}
\overline{\tilde{\Sigma}^{K}}(\omega) =
-2i\sum_{\alpha}\Gamma_{\alpha}\!\sum_{n=-\infty}^{\infty}\!\! J_{n}^{2}
\!\left(\frac{V_{ac}}{\omega_{0}}\right)\!
[1-2f(\omega-n\omega_{0}-\mu_{\alpha})].
\end{equation}

\begin{figure}
\includegraphics[width=0.98\columnwidth]{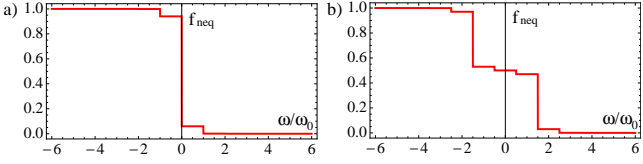}
\caption{(Color online) Distribution function of electrons in the
magnetic layer $f_{neq}(\omega)$ given by Eq.~(\ref{eq:f_neq}): (a)
pure ac voltage case, $\mu_{L}=\mu_{R}=0$ and $V_{{\rm
{ac}}}/\omega_{0}=0.5$; (b) the case of nonzero ac and dc voltages,
the parameters are $\Gamma_{L}=\Gamma_{R}$,
$\mu_{L}=-\mu_{R}=1.5 \omega_{0}$, and $V_{{\rm
{ac}}}/\omega_{0}=0.5$. Frequency $\omega$ is measured in units of
$\omega_{0}$.}
\label{fig-distr-func}
\end{figure}

The above discussion implies that the spectral function of electrons
in the magnetic layer $A_{0\sigma}= -\text{Im}[G_{0\sigma}^{R}]$ is
the same as in equilibrium, $A_{0\sigma}(\mathbf{k},\omega)=
\Gamma/[(\omega -\epsilon_{\mathbf{k}}^d -\sigma\Delta)^{2}
+\Gamma^{2}]$, where $\Gamma=\Gamma_{L}+\Gamma_{R}$ and the exchange
splitting $\Delta=J S /2$. We will assume $\Delta > \Gamma$ so that
the ferromagnetism of the conduction electrons is well defined.
Further, the nonequilibrium distribution function $f_{neq}$ of the
electrons in the magnetic layer (defined as
$G_{0\sigma}^{K}(\mathbf{k},\omega) =-2iA_{0\sigma}(\mathbf{k},\omega)
[1-2f_{neq}(\omega)]$) is
\begin{equation}
f_{neq}(\omega)=\sum_{\alpha}\frac{\Gamma_{\alpha}}{\Gamma}\sum_{n}
J_{n}^{2}\left(\frac{V_{ac}}{\omega_{0}}\right)
f(\omega -n\omega_{0}-\mu_{\alpha})
\label{eq:f_neq},
\end{equation}
Below we consider the case of zero temperature when the Fermi function
$f(\omega) =\theta(-\omega)$. A typical $f_{neq}(\omega)$ is plotted
in Fig.~\ref{fig-distr-func}. While in the pure dc case
$f_{neq}(\omega)$ is a weighted sum of Fermi functions of the left and
right leads, the ac bias adds steps to $f_{neq}$ at frequencies $\pm
|n|\omega_{0}$ corresponding to photon absorption and emission.

\section{Langevin equation} \label{secII}

We expand the effective action~(\ref{eq:eff_action}) to quadratic
order in the fluctuations to obtain,
\begin{eqnarray}
&&\!\!\!\!\!\!\!\!\!\!\!\!\!\!\!\!\!
\mathcal{S}_{K} = 2b\textrm{Tr} (S_{+}^{cl}S_{-}^{q}
+S_{+}^{q}S_{-}^{cl})
+(S_{z}^{cl} \Pi^R_{zz}S_{z}^{q}+\mathrm{H.c.})
\nonumber \\
&&\!\!\!\!\!\!\!\!\!\!\!\!\!\!\!\!\!
+ S_{z}^{q} \Pi^K_{zz}S_{z}^{q} 
+ \frac{1}{2}\!\!\!\!\sum_{\alpha,\beta =x,y} \!\!\!\!
\left[S_{\alpha}^{q} \Pi^K_{\alpha\beta}S_{\beta}^{q}
+\! (S_{\alpha}^{cl} \Pi^R_{\alpha\beta}S_{\beta}^{q} +\mathrm{H.c.})
\right]\!\! ,
\label{eq:expanded_action}
\end{eqnarray}
where $\Pi^R_{\alpha\beta}$ are the components of the polarization
operator that are calculated following standard
techniques.\cite{Mitra06} Note that
$\Pi_{xx}(\omega)=\Pi_{yy}(\omega)$ and
$\Pi_{xy}(\omega)=-\Pi_{yx}(\omega)$. Moreover, to leading order in
spin fluctuations $\Pi_{zz}$ does not play a role.  For small
frequencies $\omega\ll\Delta$, we find:
\begin{subequations}
\begin{eqnarray}
&&\!\!\!\!\!\!\!\!\!\!\Pi_{xx}^{R}(\omega)\simeq -i\beta_{xx}\omega,\quad
\Pi_{xy}^{R}(\omega)\simeq -i\beta_{xy}\omega,
\label{eq:Pi_xy_omega} \\
&&\!\!\!\!\!\!\!\!\!\!\beta_{xx}=J^{2}\nu\frac{\Gamma}{\Delta^{2}+\Gamma^{2}},
\quad\beta_{xy}=J^{2}\nu\frac{\Delta}{\Delta^{2}+\Gamma^{2}}.
\label{eq:alpha}
\end{eqnarray}
\end{subequations}
Note that $\beta_{xy}=\frac{\Delta}{\Gamma}\beta_{xx}$.

The action, Eq.~(\ref{eq:expanded_action}), may be diagonalized in the basis
$S_{\pm}^{cl}=(S_{x}^{cl}\pm iS_{y}^{cl})/2$ thus yielding the
Langevin equation \cite{Mitra06}
\begin{equation}
bS_{\pm}^{cl} +(\beta_{xx}\mp i\beta_{xy})
\dot{S}_{\pm}^{cl} = \xi_{\pm},
\label{eq:Langevin}
\end{equation}
where $\xi_{\pm} = (\xi_x \pm i\xi_y)/2$ is an auxiliary field
representing noise whose correlator is given by $\left\langle
\xi_{a=x,y}(\omega)\xi_{b=x,y}(-\omega)\right\rangle =
i\Pi_{ab}^{K}(\omega)$.  We have found an analytical expression for
$\Pi^K$ in terms of a double sum over squares of Bessel functions. For
$V_{ac}\ll\omega_{0}$ we may keep only terms corresponding to single
photon absorption and emission processes. For $\mu_L = -\mu_R = V/2$
and $\omega \ll \Delta$ the noise correlator is
\begin{eqnarray}
&&\!\!\!\!\!\!\!\!
\left\langle \xi_{x}(\omega)\xi_{x}(-\omega)\right\rangle = i \Pi^K_{xx}(\omega)\nonumber \\
&&\simeq 2\beta_{xx}\left\{
\frac{\Gamma_{L}\Gamma_{R}}{\Gamma^{2}}(\left|\omega+V\right|
+\left|\omega-V\right|)
\right.\nonumber \\
&& +\frac{\Gamma_{L}^{2}+\Gamma_{R}^{2}}{\Gamma^{2}}
\left|\omega\right|
+\left(\frac{V_{ac}}{2\omega_{0}}\right)^{2}\sum_{j=\pm 1}
\left[\frac{\Gamma_{L}^{2}
+\Gamma_{R}^{2}}{\Gamma^{2}}\left|\omega+j\omega_{0}\right|
\right.\nonumber \\
&& +\left.\left.\!\! \frac{\Gamma_{L}\Gamma_{R}}{\Gamma^{2}}
(\left|\omega+j\omega_{0}
+V \right| +\left|\omega+j\omega_{0}-V \right|)
\right]\right\}.
\label{eq:Pi_Keldysh}
\end{eqnarray}
Similarly, the off-diagonal component $\left\langle
\xi_{x}(\omega)\xi_{y}(-\omega)\right\rangle=i\Pi_{xy}^{K}(\omega)$ is
\begin{eqnarray}
\Pi_{xy}^{K}(\omega)&\!\!\simeq &\!\!
J^{2}\nu\frac{\Delta\Gamma\omega}{\left(\Delta^{2}
+\Gamma^{2}\right)^{2}}\sum_{n,m}J_{n}^{2}
\left(\frac{V_{ac}}{\omega_{0}}\right)J_{m}^{2}
\left(\frac{V_{ac}}{\omega_{0}}\right)
\nonumber \\
&\!\!\times&\!\!\!\!\!\sum_{\alpha,\beta=L,R}\!\!\!\!\frac{\Gamma_{\alpha}
\Gamma_{\beta}}{\Gamma^{2}}
\left|\omega+(m-n)\omega_{0}+\mu_{\alpha}-\mu_{\beta}\right|.
\label{eq:PiKxy}
\end{eqnarray}
An effective temperature $T_{\rm{eff}}$ may be extracted from the zero
frequency limit of the noise correlator. Eq.~(\ref{eq:Pi_Keldysh})
implies $T_{\rm{eff}} \sim (\Gamma_{L}\Gamma_{R}/\Gamma^{2})V+
(V_{ac}/2\omega_{0})^{2} [\omega_0 +
(2\Gamma_{L}\Gamma_{R}/\Gamma^{2})(V-\omega_0) \theta(V-\omega_0)]$
and therefore has a discontinuity~\cite{Lesovik94} at $V=\omega_0$. In
the opposite limit of $\omega \gg \Delta $ the noise correlator
vanishes as $\sim 1/\omega$ as expected.

Equation~(\ref{eq:Langevin}) can be rewritten in the form of a
stochastic Landau-Lifshitz-Gilbert equation~\cite{LandauLG},
\begin{equation}
\mathbf{\dot{S}} = \gamma\mathbf{S}\times\mathbf{H}_{{\rm {eff}}}
-\alpha_0 \mathbf{S}\times\mathbf{\dot{S}}+\mbox{\boldmath$\xi$}^{\prime},
\end{equation}
where $\gamma$ is the gyromagnetic ratio, $\mathbf{H}_{{\rm {eff}}}= b
\mathbf{\hat z}/(\gamma\beta_{xy})$ is an effective magnetic field,
the noise is $\mbox{\boldmath$\xi$}^{\prime} =\frac{1}{\beta_{xy}}
\hat{\mathbf{z}} \times \mbox{\boldmath$\xi$}$, and the Gilbert
damping constant is $\alpha_0 =\beta_{xx}/\beta_{xy}=\Gamma/\Delta$.

In order to determine how the magnetization dynamics affects the
$I$-$V$ characteristics we will need the spin response and correlation
functions.  From Eq.~(\ref{eq:expanded_action}) the spin-spin response
function is
\begin{equation}
D_{-+}^{R}(\omega)=i\left\langle S_{-}^{cl}(\omega)
S_{+}^{q}(-\omega)\right\rangle =\frac{1}{b-\beta_{xy}
\omega-i\beta_{xx}\omega}
\end{equation}
whereas the spin-spin correlation function is
\begin{equation}
D_{-+}^{K}(\omega)=i\left\langle S_{-}^{cl}(\omega)
S_{+}^{cl}(-\omega)\right\rangle =-\frac{(\Pi_{xx}^{K}
-i\Pi_{xy}^{K})(\omega)}{\left|b-\beta_{xy}\omega-i\beta_{xx}
\omega\right|^{2}}
\label{ss_correlator}
\end{equation}
and $D_{+-}^{R(K)}(\Delta) =D_{-+}^{R(K)}(-\Delta)$. As expected in
equilibrium ($V=V_{ac}=0$) the components of both $D(\omega)$ and
$\Pi(\omega)$ satisfy the fluctuation-dissipation theorem. 

It is instructive to take the inverse Fourier transform of 
Eq.~(\ref{ss_correlator}) to obtain the time dependence of the 
transverse spin-spin correlation function,
\begin{equation}
\left\langle S_{-}^{cl}(t>0)S_{+}^{cl}(0)\right\rangle  = 
\frac{(i\Pi_{xx}^{K}
+\Pi_{xy}^{K})(\omega_{1})}{2b\beta_{xx}}e^{-\frac{t}{\tau}-i\frac{t}{\tau_{1}}},
\label{time_ss}
\end{equation}
where $\omega_{1} = b(\beta_{xy}-i\beta_{xx})/(\beta_{xy}^{2}
+\beta_{xx}^{2})$, $\tau =\nu J^2/(b \Gamma)$, and $\tau_{1}= 2\nu
J/(bS)$. Equation~(\ref{time_ss}) shows that the spin-spin
correlations decay with the characteristic time $\tau$. Thus as long
as $1/\omega_0 \ll \tau$, the macrospin dynamics is rather slow and we
can use the time-averaging procedure for the Green's functions
outlined in Sec.~\ref{secI}.

\section{Current-voltage characteristics} \label{secIII}

\begin{figure}
\includegraphics[width=0.98\columnwidth]{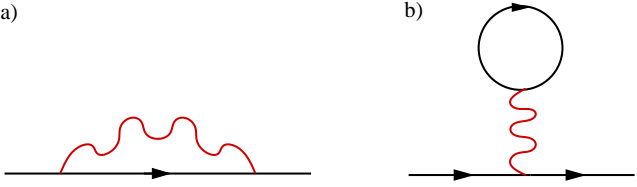}
\caption{(Color online) The diagrams for one-loop corrections to the
electron Green's function in the magnetic layer due to coupling to
spin fluctuations: (a) exchange (Fock) contribution $\Sigma_{\sigma
d}^{eR}$, (b) Hartree contribution $\Sigma_{\sigma d}^{hR}$. Wavy
lines correspond to spin-spin correlators $\left\langle
S_{-}(\omega)S_{+}(-\omega)\right\rangle $.}
\label{fig-diagrams}
\end{figure}

We will now study how the current-voltage characteristics of the
magnetic junction are affected by the magnetization dynamics of the
ferromagnetic layer. We employ the Jauho-Meir-Wingreen
formula \cite{Wingreen} for the tunneling current
\begin{equation}
I =\frac{e}{\hbar}\int\frac{d\Omega}{2\pi}\sum_{\mathbf{k}, \sigma}
[f(\Omega-\mu_{L})-f(\Omega-\mu_{R})]
\frac{4\Gamma_{L}\Gamma_{R}}{\Gamma}A(\mathbf{k},\Omega).
\label{eq:tunnel_Current}
\end{equation}
In the following we calculate the leading correction to the spectral
function $A_{\sigma}= - \text{Im}[G_{\sigma d}^{R}]$ due to coupling
to spin fluctuations.  The spectral function is determined from the
Dyson equation $\left[G_{\sigma d}^{R}\right]^{-1}=\left[G_{\sigma
0}^{R}\right]^{-1}- \Sigma_{\sigma d}^R$, where $\Sigma_{\sigma d}$ is
the self-energy due to coupling to spin fluctuations.  To one-loop
order $\Sigma_{\sigma d}^{R} = \Sigma_{\sigma d}^{eR} +\Sigma_{\sigma
d}^{h R}$, where $\Sigma_{\sigma d}^{eR}$ and $\Sigma_{\sigma d}^{hR}$
are, respectively, the exchange and Hartree contributions to the
self-energy, see Fig.~\ref{fig-diagrams}.  To leading order in the
fluctuations, it suffices to do perturbation theory in $J^2$ so that
$G_{d\sigma}^R = G_{0\sigma}^R + \delta G_{d\sigma}^R$ with
\begin{equation}
\delta G_{d\sigma}^{R} =G_{0\sigma}^{R}\Sigma_{\sigma d}^{R}
G_{0\sigma}^{R}.
\label{eq:delta_G_R}
\end{equation}
Note that $\Sigma_{\uparrow d}$ and $\Sigma_{\downarrow d}$ are
related by $\Delta \leftrightarrow -\Delta$.

The exchange contribution to the self-energy is
\begin{eqnarray}
\Sigma_{\downarrow d}^{eR}(\mathbf{k},\Omega) &=&
-\frac{iJ^{2}}{2}\!\int\!\frac{d\omega}{2\pi}\!\!
\left[G_{d\uparrow}^{R}(\mathbf{k},\omega+\Omega)D_{-+}^{K}(\omega)
  \right.  \nonumber \\
&&\left. +G_{d\uparrow}^{K}(\mathbf{k},\omega+\Omega)D_{-+}^{A}(\omega)\right].
\label{eq:self_energy_up}
\end{eqnarray}
Keeping terms to leading order in $J^{2}$, $\Sigma^{eR}$ is purely
real and given by
\begin{eqnarray}
&&\Sigma_{\downarrow d}^{eR}(\mathbf{k},\Omega) =
-\frac{J^{2}}{\pi b}\sum_{\alpha}
\frac{\Gamma_{\alpha}}{\Gamma}\left[\frac{\pi}{2}
+\arctan\frac{\epsilon_{\mathbf{k}}^d
+\Delta-\mu_{\alpha}}{\Gamma} \right.\nonumber \\
&&+\!\! \left.\left(\frac{V_{ac}}{2\omega_{0}}\right)^{2}\!
\left(\pi+\sum_{m =\pm 1} \arctan\frac{\epsilon_{\mathbf{k}}^d+\Delta
+m\omega_{0}-\mu_{\alpha}}{\Gamma} \right) \right]\! .\nonumber \\
\label{eq:Sige1}
\end{eqnarray}
The Hartree contribution to the self-energy is given by
\begin{equation}
\Sigma_{\downarrow d}^{hR}(\mathbf{k},\Omega)
= -\frac{iJ^{2}}{2}D_{+-}^{A}(\omega=0)
\int\nu d\epsilon'\int \frac{d\omega^{\prime}}{2\pi}
G_{d\uparrow}^{K}(\epsilon^{\prime},\omega^{\prime})\nonumber
\end{equation}
where $D_{+-}^{A}(\omega=0) = 1/b$ and we have set
$\sum_{\mathbf{k}^{\prime}} \to\int\nu d\epsilon^{\prime}$. Note that
the Hartree contribution is independent of external frequency and
momentum, and therefore only shifts the position of the pole of
$G_{d}^{R}$ but does not contribute to the corrections to the current.

We denote the total current averaged over time $T \gg 1/\omega_0$ as
$\bar{I}= I_0 + \delta \bar{I}$ where
\begin{equation}
I_0= 4
\frac{e}{\hbar}\frac{\nu\Gamma_{L}\Gamma_{R} }{\Gamma}V 
\end{equation}
is the current for a static ferromagnet while $\delta \bar{I}$ is the
leading correction due to spin fluctuations computed from
Eqs.~(\ref{eq:tunnel_Current}),~(\ref{eq:delta_G_R}),
and~(\ref{eq:Sige1}) for $\mu_L = -\mu_R = V/2$
\begin{eqnarray}
\frac{\delta \bar{I}}{I_0}\!\! &\simeq &\!\! \frac{\Gamma J^{2}}
{2 \pi b\Delta^{2}\left( 1 +\frac{\Gamma^{2}}{\Delta^2}\right)}\left\{\!\!
\left(1+\frac{V_{ac}^2}{2\omega_{0}^2}\right)\!\!
\left[1+\frac{(3\Delta^{2}-\Gamma^{2})V^2}{12(\Delta^{2}
+\Gamma^{2})^{2}} \right]\right.\nonumber \\
&&+ \left.\!\frac{V_{ac}^2}{16}\left[
\frac{3\Delta^{2}-\Gamma^{2}}{(\Delta^{2}+\Gamma^{2})^{2}}
+\frac{(\sqrt{5}\Delta^{2}-\Gamma^{2})^{2}V^{2}}{2
\left(\Delta^{2}+\Gamma^{2}\right)^{4}}\right]\right\}.
\label{eq:current_total}
\end{eqnarray}
Since $\Delta \sim JS$, this correction to the current is ${\cal
  O}(\Gamma /bS^{2})$.  Thus our perturbative treatment in spin
fluctuations is valid as long as $b\neq 0$ and $S\gg 1$.  Moreover the
correction $\delta \bar{I} > 0$. This is because scattering off spin
fluctuations in this geometry produces additional channels for
electron conduction (in contrast to a bulk geometry where this
scattering would cause the conductivity to decrease).  It is worth
mentioning that ac bias contributes to the Ohmic corrections as well
with terms such as $\sim (V_{ac}^{2}/\Delta^2)V$ and
$(V_{ac}^{2}/\omega_{0}^{2})V$. For the pure dc case ($V_{ac}=0$), we
find that the differential conductance $g =e \partial I/\partial V$
for $\Delta \gg\Gamma$ is
\begin{equation}
g = 4 \frac{e^2}{\hbar}\frac{\nu\Gamma_{L}\Gamma_{R}
}{\Gamma} \left[1+ \frac{J^{2} \Gamma}{2\pi b
\Delta^{2}}\left(1+\frac{3V^2}{4\Delta^{2}} \right)\right].
\label{conductance}
\end{equation}

Note that the quadratic in voltage corrections in
Eq. (\ref{conductance}) are similar in spirit to temperature
corrections ($\sim T^2$) to the conductance. Also, our result is for a
particular choice of chemical potentials $\mu_L = -\mu_R = V/2$. The
answer in general will change if a different choice, such as $\mu_L =
V$ and $\mu_R=0$, were used.  The reason for this difference is that
the symmetric combination of chemical potentials $(\mu_L + \mu_R)/2$
plays the role of a mean chemical potential for the electrons in the
nanomagnet, tuning which modifies the equilibrium spectral density and
hence the linear-response conductance as well as other equilibrium
properties. In an experiment this mean chemical potential may be tuned
by an external gate voltage.  Our purely antisymmetric combination
$\mu_L = -\mu_R$ avoids these intrinsically equilibrium effects.

\section{Summary} \label{secIV}

We have derived a Langevin equation for the magnetization dynamics for
a simultaneously applied ac and dc bias across an N/F/N
nanostructure. The magnetization dynamics is characterized by a
frequency dependent noise, Eq.~(\ref{eq:Pi_Keldysh}).  We have also
computed corrections to the $I$-$V$ characteristics to leading
($1/S^2$) order in the spin-fluctuations. These fluctuations are found
to not only modify the Ohmic part of the $I$-$V$ characteristics, but
to also give rise to corrections that are non-linear in voltage,
Eq.~(\ref{eq:current_total}). Experiments often exhibit non-linear
$I$-$V$ curves,\cite{magnet_dyn_exp} but the origin of the
nonlinearities is usually not clear. The usefulness of our result is
that the current is a function of three independent experimentally
tunable parameters (the dc bias $V$, the ac amplitude $V_{ac}$ and
frequency $\omega_0$) which can in principle allow one to extract the
physics arising only from magnetization dynamics.

\acknowledgments

We are grateful to A.~D.~Kent and D.~Bedau for valuable discussions.
This work was supported by the NSF-DMR (Grant No. 0705584).


\begin{thebibliography}{99}
\bibitem{1} This is under the assumption that the nanomagnet is
  strongly coupled to the leads so that the escape time of electrons
  into the leads is much shorter than the dwell time.
\end{thebibliography}
\end{document}